# PROCA EQUATION
# FOR ATTOSECOND ELECTRON PULSES

[1]Magdalena Pelc, [2]Janina Marciak-Kozlowska, [3*] Miroslaw Kozlowski

[1]Institute of Physics Marie Curie-Sklodowska University, Lublin, Poland
[2] Institute of Electron Technology , Al. Lotnikow 32/46,02-668 Warsaw Poland
[3*] Institute of Experimental Physics, Warszaw University
Hoza 69 Warszaw , Poland

* Corresponding author

Abstract

In this paper the heat transport of attosecond electron pulses is investigated.
It is shown that attosecond electrons can propagate as thermal waves or diffused as particle conglommerates ,Proca equation *as* type equation for the thermal transport of the attosecond electron pulsem is formulated

Key words: attosecond electron pulses, heat transport, *Proca* equation

## 1 Introduction

In the seminal paper [1] prof Ahmed Zewail and P Baum put forward the very interesting idea of the creation of the attosecond electron pulses . In this paper we develope the theoretical model for the heat transport of the attosecond electron pulses . As can be shown the electron heated by attosecond electron pulses can evolve as the heat wave, due to hyperbolicity of the master equation

Parabolic theories of dissipative phenomena have long and a venerable history and proved very useful, especially in the steady-state regime [2]. They exhibit, however, some undesirable features, such as a causality (see e.g., [2]), which prompted the formulation of hyperbolic theories of dissipation to get rid of them. This was achieved at a price by extending the set of field variables by including the dissipative fluxes (heat current, non-equilibrium stresses and so on) on the same footing as the classical (energy densities, equilibrium pressures, etc), thereby giving rise to a set of 'more physically' satisfactory (as they conform much better with experiments) but involved theories from the mathematical point of view. These theories have the additional advantage of being backed by statistical fluctuation theory, the kinetic theory of gases (Grad's 13-moment approximation), information theory and correlated random walks

A key quantity in these theories is the relaxation time $\tau$ of the corresponding dissipative process. This positive-definite quantity has a distinct physical meaning, namely the time taken by the system to return spontaneously to the steady state (whether of thermodynamic equilibrium or not) after it has been suddenly removed from it. It is, however, connected to the mean collision time $t_c$ of the particles responsible for the dissipative process It is therefore appropriate to interpret the relaxation time as the time taken by the corresponding dissipative flow to relax to its steady value.

In the book [2] the new hyperbolic Proca type equation for heat transport was formulated and solved.

The excitation of matter on the atomic level leads to transfer of energy. The response of the matter is governed by the relaxation time.

In this paper we develop the general, universal definition of the relaxation time, which depends on coupling constants for electromagnetic interaction.

It occurs that the general formula for the relaxation time can be written as

$$\tau_i = \frac{\hbar}{m_i(\alpha_i c)^2} \tag{1}$$

where $m_i$ is the heat carrier mass, $\alpha_i = \left(i = e, \tfrac{1}{137}\right)$ is coupling constant for electromagnetic interaction, $c$ is the vacuum light speed. As the $c$ is the maximum speed all relaxation time fulfils the inequality

$$\tau > \tau_i$$

Consequently $\tau_i$ is the minimal universal relaxation time.

## 2 Proca heat transport equation

After the standards of time and space were defined, the laws of classical physics relating such parameters as distance, time, velocity, temperature are assumed to be independent of the accuracy with which, these parameters can be measured. It should be noted that this assumption does not enter explicitly into the formulation of classical physics. It implies that together with the assumption of existence of an object and really independently of any measurements (in classical physics) it was tacitly assumed that *there was a possibility of an unlimited increase in the accuracy of measurements.* Bearing in mind the "atomicity" of time i.e. considering the smallest time period, the Planck time, the above statement is obviously not true. Atto-second electron pulses are at the limit of time resolution.

In this paragraph, we develop and solve the quantum relativistic heat transport equation for atto-second electron transport phenomena, where external forces exist [5]. In paragraph 2 we develop the new hyperbolic heat transport equation which generalises the Fourier heat transport equation for the rapid thermal processes has been written in the form (3):

$$\frac{1}{\left(\frac{1}{3}\upsilon_F^2\right)}\frac{\partial^2 T}{\partial t^2}+\frac{1}{\tau\left(\frac{1}{3}\upsilon_F^2\right)}\frac{\partial T}{\partial t}=\nabla^2 T\ , \tag{3}$$

where $T$ denotes the temperature, $\tau$ the relaxation time for the thermal disturbance of the fermionic system, and $\upsilon_F$ is the Fermi velocity.

In what follows we present the formulation of the HHT, considering the details of the two fermionic systems: electron gas in metals .

For the electron gas in metals, the Fermi energy has the form

$$E_F^e=(3\pi)^2\frac{n^{2/3}\hbar^2}{2m_e}, \tag{4}$$

where $n$ denotes the density and $m_e$ electron mass. Considering that

$$n^{-1/3}\sim a_B\sim\frac{\hbar^2}{me^2}, \tag{5}$$

and $a_B$ = Bohr radius, one obtains

$$E_F^e\sim\frac{n^{2/3}\hbar^2}{2m_e}\sim\frac{\hbar^2}{ma^2}\sim\alpha^2 m_e c^2, \tag{6}$$

where $c$ = light velocity and $\alpha$ = 1/137 is the fine-structure constant for electromagnetic interaction. For the Fermi momentum $p_F$ we have

$$p_F^e\sim\frac{\hbar}{a_B}\sim\alpha m_e c, \tag{7}$$

and, for Fermi velocity $\upsilon_F$ ,

$$\upsilon_F^e\sim\frac{p_F}{m_e}\sim\alpha c. \tag{8}$$

Equation (8) gives the theoretical background for the result presented in paragraph 2. Considering formula (8), equation HHT can be written as

$$\frac{1}{c^2}\frac{\partial^2 T}{\partial t^2}+\frac{1}{c^2\tau}\frac{\partial T}{\partial t}=\frac{\alpha^2}{3}\nabla^2 T. \tag{9}$$

As seen from (9), the HHT equation is a relativistic equation, since it takes into account the finite velocity of light.

In the following, the procedure for the quantisation of temperature $T(\vec{r},t)$ in a hot fermion gas will be developed. First of all, we introduce the reduced de Broglie wavelength

$$\lambda_B^e = \frac{\hbar}{m_e \upsilon_h^e}, \qquad \upsilon_h^e = \frac{1}{\sqrt{3}}\alpha c, \tag{10}$$

and the mean free path $\lambda^e$

$$\lambda^e = \upsilon_h^e \tau^e, \tag{11}$$

In view of the equations (10) and (11), we obtain the HHC for electron and nucleon gases

$$\frac{\lambda_B^e}{\upsilon_h^e}\frac{\partial^2 T}{\partial t^2} + \frac{\lambda_B^e}{\lambda^e}\frac{\partial T}{\partial t} = \frac{\hbar}{m_e}\nabla^2 T^e, \tag{12}$$

Equations (11) and (12) are the hyperbolic partial differential equations which are the master equations for heat propagation in Fermi electron and nucleon gases. In the following, we study the quantum limit of heat transport in the fermionic systems. We define the quantum heat transport limit as follows:

$$\lambda^e = \lambda_B^e, \tag{13}$$

In that case, Eq. (11) has the form

$$\tau^e \frac{\partial^2 T^e}{\partial t^2} + \frac{\partial T^e}{\partial t} = \frac{\hbar}{m_e}\nabla^2 T^e, \tag{14}$$

where

$$\tau^e = \frac{\hbar}{m_e\left(\upsilon_h^e\right)^2}, \qquad (15)$$

Equations (14) and (15) define the master equation for quantum heat transport (QHT). With the relaxation times $\tau^e$ one can define the "pulsations" $\omega_h^e$

$$\omega_h^e = (\tau^e)^{-1}, \qquad (16)$$

or

$$\omega_h^e = \frac{m_e\left(\upsilon_h^e\right)^2}{\hbar},$$

i.e.,

$$\omega_h^e \hbar = m_e\left(\upsilon_h^e\right)^2 = \frac{m_e \alpha^2}{3} c^2, \qquad (17)$$

The equations (17) define the Planck-Einstein relation for heat quanta $E_h^e$ and $E_h^N$

$$E_h^e = \omega_h^e \hbar = m_e\left(\upsilon_h^e\right)^2, \qquad (18)$$

The heat quantum with the energy $E_h = \hbar\omega$ can be named the *heaton*, in complete analogy to the *phonon, magnon, roton,* etc. For $\tau^e$ ,→ 0, Eq. (14) are the Fourier equations with quantum diffusion coefficients $D^e$ and

$$\frac{\partial T^e}{\partial t} = D^e \nabla^2 T^e, \qquad D^e = \frac{\hbar}{m_e}, \qquad (19)$$

For finite $\tau^e$, for $\Delta t < \tau^e$, , Eq. (14) can be written as

$$\frac{1}{(\upsilon_h^e)^2}\frac{\partial^2 T^e}{\partial t^2}=\nabla^2 T^e, \tag{20}$$

Equations (19) and (20) are the wave equations for quantum heat transport (QHT)

It is quite interesting that the Proca type equation can be obtained for thermal phenomena. In the following, starting with the hyperbolic heat diffusion equation the Proca equation for thermal processes will be developed [2]. (21)

When the external force is present *F(x,t)* the forced damped heat transport is obtained [2] (in a one dimensional case):

$$\frac{1}{\upsilon^2}\frac{\partial^2 T}{\partial t^2}+\frac{m_0\gamma}{\hbar}\frac{\partial T}{\partial t}+\frac{2Vm_0\gamma}{\hbar^2}T-\frac{\partial^2 T}{\partial x^2}=F(x,t). \tag{22}$$

The hyperbolic relativistic quantum heat transport equation, (22), describes the forced motion of heat carriers, which undergo scattering ($\frac{m_0\gamma}{\hbar}\frac{\partial T}{\partial t}$ term) and are influenced by the potential term ($\frac{2Vm_o\gamma}{\hbar^2}T$).

Equation (22) is the Proca thermal equation and can be written as [2]:

$$\begin{gathered}\left(\bar{\Box}^2+\frac{2Vm_0\gamma}{\hbar^2}\right)T+\frac{m_0\gamma}{\hbar}\frac{\partial T}{\partial t}=F(x,t),\\ \bar{\Box}^2=\frac{1}{\upsilon^2}\frac{\partial^2}{\partial t^2}-\frac{\partial^2}{\partial x^2}.\end{gathered} \tag{24}$$

We seek the solution of equation (24) in the form

$$T(x,t)=e^{-t/2\tau}u(x,t), \tag{25}$$

where $\tau_i=\frac{\hbar}{m\upsilon^2}$ is the relaxation time. After substituting equation (25) in equation (24) we obtain a new equation

$$\left(\bar{\Box}^2+q\right)u(x,t)=e^{t/2\tau}F(x,t) \tag{26}$$

and

$$q = \frac{2Vm}{\hbar^2} - \left(\frac{m\upsilon}{2\hbar}\right)^2, \tag{27}$$

$$m = m_0\gamma \ . \tag{28}$$

In free space i.e. when $F(x,t) \to 0$ equation (24) reduces to

$$\left(\Box^2 + q\right)u(x,t) = 0\,, \tag{42}$$

which is essentially the free Proca type equation.

The Proca equation describes the interaction of the atto-second electron pulse with the matter. As was shown in book [2] the quantization of the temperature field leads to the *heatons* – quanta of thermal energy with a mass $m_h = \frac{\hbar}{\tau\upsilon_h^2}$ [2], where $\tau$ is the relaxation time and $\upsilon_h$ is the finite velocity for heat propagation. With $\upsilon_h \to \infty$, i.e. for $c \to \infty$, $m_0 \to 0$, it can be concluded that in non-relativistic approximation ($c$ = infinite) the *Proca* equation is the diffusion equation for mass-less photons and heatons.

## 3 Conclusion

In this paper the *Proca* thermal equation ( 26} was formulated and solved. The proca equation is the hyperbolic partial differential equation. The solution of the Proca equation the damped thermal waves can propagate in electron gas with the speed v=α c, where α=1/137 is the fine structure constant and c= vacuum light velocity